\documentclass[article,aps,nofootinbib,twocolumn,superscriptaddress]{revtex4}
\input epsf
\usepackage{graphics}
\usepackage{amsmath}
\usepackage{color}
\usepackage{dcolumn}
\usepackage{hyphenat}

\usepackage{graphicx}

\def\be{\begin{equation}}
\def\ee{\end{equation}}
\def\ba{\begin{eqnarray}}
\def\ea{\end{eqnarray}}

\newcommand{\beqa}{\begin{eqnarray}}
\newcommand{\eeqa}{\end{eqnarray}}

\newcommand{\beq}{\begin{equation}}
\newcommand{\eeq}{\end{equation}}
\newcommand{\bfl}{{\mathbf{l}}}

\newlength{\tskip}
\newlength{\colwidth}

\newcommand{\Pb}{\textsc{Polarbear}}
\newcommand{\pb}{\textsc{\mbox{Polarbear}}}

\newcommand{\pmfupper}{3.9}

\begin{document}

\title{POLARBEAR Constraints on Cosmic Birefringence and Primordial Magnetic Fields}

\author{\Pb\ Collaboration} \noaffiliation
\author{Peter A.R. Ade} \affiliation{School of Physics and Astronomy, Cardiff University, Cardiff CF10 3XQ, United Kingdom}
\author{Kam Arnold} \affiliation{Department of Physics, University of California, San Diego, CA 92093-0424, USA}
\author{Matt Atlas} \affiliation{Department of Physics, University of California, San Diego, CA 92093-0424, USA}
\author{Carlo Baccigalupi} \affiliation{International School for Advanced Studies (SISSA), Via Bonomea 265, 34136, Trieste, Italy}
\author{Darcy Barron} \affiliation{Department of Physics, University of California, Berkeley, CA 94720, USA}
\author{David Boettger} \affiliation{Department of Astronomy, Pontifica Universidad Catolica, Santiago, Chile}
\author{Julian Borrill} \affiliation{Computational Cosmology Center, Lawrence Berkeley National Laboratory, Berkeley, CA 94720, USA}\affiliation{Space Sciences Laboratory, University of California, Berkeley, CA 94720, USA}
\author{Scott Chapman} \affiliation{Department of Physics and Atmospheric Science, Dalhousie University, Halifax, NS, B3H 4R2, Canada}
\author{Yuji Chinone} \affiliation{Department of Physics, University of California, Berkeley, CA 94720, USA}
\author{Ari Cukierman} \affiliation{Department of Physics, University of California, Berkeley, CA 94720, USA}
\author{Matt Dobbs} \affiliation{Physics Department, McGill University, Montreal, QC H3A 0G4, Canada}
\author{Anne Ducout} \affiliation{Department of Physics, Blackett Laboratory, Imperial College London, London SW7 2AZ, United Kingdom}
\author{Rolando Dunner} \affiliation{Department of Astronomy, Pontifica Universidad Catolica, Santiago, Chile}
\author{Tucker Elleflot} \affiliation{Department of Physics, University of California, San Diego, CA 92093-0424, USA}
\author{Josquin Errard} \affiliation{Space Sciences Laboratory, University of California, Berkeley, CA 94720, USA}\affiliation{Computational Cosmology Center, Lawrence Berkeley National Laboratory, Berkeley, CA 94720, USA}
\author{Giulio Fabbian} \affiliation{International School for Advanced Studies (SISSA), Via Bonomea 265, 34136, Trieste, Italy}
\author{Stephen Feeney} \affiliation{Department of Physics, Blackett Laboratory, Imperial College London, London SW7 2AZ, United Kingdom}
\author{Chang Feng} \affiliation{Department of Physics and Astronomy, University of California, Irvine, CA 92697-4575, USA}
\author{Adam Gilbert} \affiliation{Physics Department, McGill University, Montreal, QC H3A 0G4, Canada}
\author{Neil Goeckner-Wald} \affiliation{Department of Physics, University of California, Berkeley, CA 94720, USA}
\author{John Groh} \affiliation{Department of Physics, University of California, Berkeley, CA 94720, USA}
\author{Grantland Hall} \affiliation{Department of Physics, University of California, Berkeley, CA 94720, USA}
\author{Nils W. Halverson} \affiliation{Center for Astrophysics and Space Astronomy, University of Colorado, Boulder, CO 80309, USA}\affiliation{Department of Astrophysical and Planetary Sciences, University of Colorado, Boulder, CO 80309, USA}\affiliation{Department of Physics, University of Colorado, Boulder, CO 80309, USA}
\author{Masaya Hasegawa} \affiliation{High Energy Accelerator Research Organization (KEK), Tsukuba, Ibaraki 305-0801, Japan}\affiliation{SOKENDAI (The Graduate University for Advanced Studies), Hayama, Miura District, Kanagawa 240-0115, Japan}
\author{Kaori Hattori} \affiliation{High Energy Accelerator Research Organization (KEK), Tsukuba, Ibaraki 305-0801, Japan}
\author{Masashi Hazumi} \affiliation{High Energy Accelerator Research Organization (KEK), Tsukuba, Ibaraki 305-0801, Japan}\affiliation{Kavli IPMU (WPI), UTIAS, The University of Tokyo, Kashiwa, Chiba 277-8583, Japan}\affiliation{SOKENDAI (The Graduate University for Advanced Studies), Hayama, Miura District, Kanagawa 240-0115, Japan}
\author{Charles Hill} \affiliation{Department of Physics, University of California, Berkeley, CA 94720, USA}
\author{William L. Holzapfel} \affiliation{Department of Physics, University of California, Berkeley, CA 94720, USA}
\author{Yasuto Hori} \affiliation{Department of Physics, University of California, Berkeley, CA 94720, USA}
\author{Logan Howe} \affiliation{Department of Physics, University of California, San Diego, CA 92093-0424, USA}
\author{Yuki Inoue} \affiliation{SOKENDAI (The Graduate University for Advanced Studies), Hayama, Miura District, Kanagawa 240-0115, Japan}\affiliation{High Energy Accelerator Research Organization (KEK), Tsukuba, Ibaraki 305-0801, Japan}
\author{Gregory C. Jaehnig} \affiliation{Center for Astrophysics and Space Astronomy, University of Colorado, Boulder, CO 80309, USA}\affiliation{Department of Physics, University of Colorado, Boulder, CO 80309, USA}
\author{Andrew H. Jaffe} \affiliation{Department of Physics, Blackett Laboratory, Imperial College London, London SW7 2AZ, United Kingdom}
\author{Oliver Jeong} \affiliation{Department of Physics, University of California, Berkeley, CA 94720, USA}
\author{Nobuhiko Katayama} \affiliation{Kavli IPMU (WPI), UTIAS, The University of Tokyo, Kashiwa, Chiba 277-8583, Japan}
\author{Jonathan P. Kaufman} \affiliation{Department of Physics, University of California, San Diego, CA 92093-0424, USA}
\author{Brian Keating} \affiliation{Department of Physics, University of California, San Diego, CA 92093-0424, USA}
\author{Zigmund Kermish} \affiliation{Department of Physics, Princeton University, Princeton, NJ 08544, USA}
\author{Reijo Keskitalo} \affiliation{Computational Cosmology Center, Lawrence Berkeley National Laboratory, Berkeley, CA 94720, USA}\affiliation{Space Sciences Laboratory, University of California, Berkeley, CA 94720, USA}
\author{Theodore Kisner} \affiliation{Computational Cosmology Center, Lawrence Berkeley National Laboratory, Berkeley, CA 94720, USA}\affiliation{Space Sciences Laboratory, University of California, Berkeley, CA 94720, USA}
\author{Akito Kusaka} \affiliation{Physics Division, Lawrence Berkeley National Laboratory, Berkeley, CA 94720, USA}
\author{Maude Le Jeune} \affiliation{AstroParticule et Cosmologie, Univ Paris Diderot, CNRS/IN2P3, CEA/Irfu, Obs de Paris, Sorbonne Paris Cit\'e, France, }
\author{Adrian T. Lee} \affiliation{Department of Physics, University of California, Berkeley, CA 94720, USA}\affiliation{Physics Division, Lawrence Berkeley National Laboratory, Berkeley, CA 94720, USA}
\author{Erik M. Leitch} \affiliation{Department of Astronomy and Astrophysics, University of Chicago, Chicago, IL 60637, USA}\affiliation{Kavli Institute for Cosmological Physics, University of Chicago, Chicago, IL 60637, USA}
\author{David Leon} \affiliation{Department of Physics, University of California, San Diego, CA 92093-0424, USA}
\author{Yun Li} \affiliation{Department of Physics, Simon Fraser University, Burnaby, BC, V5A 1S6, Canada}
\author{Eric Linder} \affiliation{Physics Division, Lawrence Berkeley National Laboratory, Berkeley, CA 94720, USA}
\author{Lindsay Lowry} \affiliation{Department of Physics, University of California, San Diego, CA 92093-0424, USA}
\author{Frederick Matsuda} \affiliation{Department of Physics, University of California, San Diego, CA 92093-0424, USA}
\author{Tomotake Matsumura} \affiliation{Institute of Space and Astronautical Studies (ISAS), Japan Aerospace Exploration Agency (JAXA), Sagamihara, Kanagawa 252-5210, Japan}
\author{Nathan Miller} \affiliation{Observational Cosmology Laboratory, Code 665, NASA Goddard Space Flight Center, Greenbelt, MD 20771, USA}
\author{Josh Montgomery} \affiliation{Physics Department, McGill University, Montreal, QC H3A 0G4, Canada}
\author{Michael J. Myers} \affiliation{Department of Physics, University of California, Berkeley, CA 94720, USA}
\author{Martin Navaroli} \affiliation{Department of Physics, University of California, San Diego, CA 92093-0424, USA}
\author{Haruki Nishino} \affiliation{High Energy Accelerator Research Organization (KEK), Tsukuba, Ibaraki 305-0801, Japan}
\author{Takahiro Okamura} \affiliation{High Energy Accelerator Research Organization (KEK), Tsukuba, Ibaraki 305-0801, Japan}
\author{Hans Paar} \affiliation{Department of Physics, University of California, San Diego, CA 92093-0424, USA}
\author{Julien Peloton} \affiliation{AstroParticule et Cosmologie, Univ Paris Diderot, CNRS/IN2P3, CEA/Irfu, Obs de Paris, Sorbonne Paris Cit\'e, France, }
\author{Levon Pogosian} \affiliation{Department of Physics, Simon Fraser University, Burnaby, BC, V5A 1S6, Canada}
\author{Davide Poletti} \affiliation{AstroParticule et Cosmologie, Univ Paris Diderot, CNRS/IN2P3, CEA/Irfu, Obs de Paris, Sorbonne Paris Cit\'e, France, }
\author{Giuseppe Puglisi} \affiliation{International School for Advanced Studies (SISSA), Via Bonomea 265, 34136, Trieste, Italy}
\author{Christopher Raum} \affiliation{Department of Physics, University of California, Berkeley, CA 94720, USA}
\author{Gabriel Rebeiz} \affiliation{Department of Electrical and Computer Engineering, University of California, San Diego, CA 92093, USA}
\author{Christian L. Reichardt} \affiliation{School of Physics, University of Melbourne, Parkville, VIC 3010, Australia}
\author{Paul L. Richards} \affiliation{Department of Physics, University of California, Berkeley, CA 94720, USA}
\author{Colin Ross} \affiliation{Department of Physics and Atmospheric Science, Dalhousie University, Halifax, NS, B3H 4R2, Canada}
\author{Kaja M. Rotermund} \affiliation{Department of Physics and Atmospheric Science, Dalhousie University, Halifax, NS, B3H 4R2, Canada}
\author{David E. Schenck} \affiliation{Center for Astrophysics and Space Astronomy, University of Colorado, Boulder, CO 80309, USA}\affiliation{Department of Astrophysical and Planetary Sciences, University of Colorado, Boulder, CO 80309, USA}
\author{Blake D. Sherwin} \affiliation{Department of Physics, University of California, Berkeley, CA 94720, USA}\affiliation{Miller Institute for Basic Research in Science, University of California, Berkeley, CA 94720, USA}
\author{Meir Shimon} \affiliation{School of Physics and Astronomy, Tel Aviv University, Tel Aviv 69978, Israel}
\author{Ian Shirley} \affiliation{Department of Physics, University of California, Berkeley, CA 94720, USA}
\author{Praween Siritanasak} \affiliation{Department of Physics, University of California, San Diego, CA 92093-0424, USA}
\author{Graeme Smecher} \affiliation{Physics Department, McGill University, Montreal, QC H3A 0G4, Canada}
\author{Nathan Stebor} \affiliation{Department of Physics, University of California, San Diego, CA 92093-0424, USA}
\author{Bryan Steinbach} \affiliation{Department of Physics, University of California, Berkeley, CA 94720, USA}
\author{Aritoki Suzuki} \affiliation{Radio Astronomy Laboratory, University of California, Berkeley, CA 94720, USA}
\author{Jun-ichi Suzuki} \affiliation{High Energy Accelerator Research Organization (KEK), Tsukuba, Ibaraki 305-0801, Japan}
\author{Osamu Tajima} \affiliation{High Energy Accelerator Research Organization (KEK), Tsukuba, Ibaraki 305-0801, Japan}\affiliation{SOKENDAI (The Graduate University for Advanced Studies), Hayama, Miura District, Kanagawa 240-0115, Japan}
\author{Satoru Takakura} \affiliation{Osaka University, Toyonaka, Osaka 560-0043, Japan}\affiliation{High Energy Accelerator Research Organization (KEK), Tsukuba, Ibaraki 305-0801, Japan}
\author{Alexei Tikhomirov} \affiliation{Department of Physics and Atmospheric Science, Dalhousie University, Halifax, NS, B3H 4R2, Canada}
\author{Takayuki Tomaru} \affiliation{High Energy Accelerator Research Organization (KEK), Tsukuba, Ibaraki 305-0801, Japan}
\author{Nathan Whitehorn} \affiliation{Department of Physics, University of California, Berkeley, CA 94720, USA}
\author{Brandon Wilson} \affiliation{Department of Physics, University of California, San Diego, CA 92093-0424, USA}
\author{Amit Yadav} \affiliation{Department of Physics, University of California, San Diego, CA 92093-0424, USA}
\author{Alex Zahn} \affiliation{Department of Physics, University of California, San Diego, CA 92093-0424, USA}
\author{Oliver Zahn} \affiliation{Department of Physics, University of California, Berkeley, CA 94720, USA}

\begin{abstract}
We constrain anisotropic cosmic birefringence using four-point correlations of even-parity $E$-mode and odd-parity $B$-mode polarization in the cosmic microwave background measurements made by the POLARization of the Background Radiation (\pb) experiment in its first season of observations. We find that the anisotropic cosmic birefringence signal from any parity-violating processes is consistent with zero. The Faraday rotation from anisotropic cosmic birefringence can be compared with the equivalent quantity generated by primordial magnetic fields if they existed. The \pb\ nondetection translates into a 95\% confidence level (C.L.) upper limit of 93 nanogauss (nG) on the amplitude of an equivalent primordial magnetic field inclusive of systematic uncertainties. This four-point correlation constraint on Faraday rotation is about 15 times tighter than the upper limit of 1380 nG inferred from constraining the contribution of Faraday rotation to two-point correlations of $B$-modes measured by Planck in 2015. Metric perturbations sourced by primordial magnetic fields would also contribute to the $B$-mode power spectrum. Using the \pb\ measurements of the $B$-mode power spectrum (two-point correlation), we set a 95\% C.L. upper limit of \pmfupper\ nG on primordial magnetic fields assuming a flat prior on the field amplitude. This limit is comparable to what was found in the Planck 2015 two-point correlation analysis with both temperature and polarization. We perform a set of systematic error tests and find no evidence for contamination. This work marks the first time that anisotropic cosmic birefringence or primordial magnetic fields have been constrained from the ground at subdegree scales.
\end{abstract}

\maketitle

\section{Introduction}

The cosmic microwave background (CMB) has been an invaluable resource for testing fundamental physics. The CMB anisotropy has a polarized component that can be separated into even-parity ($E$-mode) and odd-parity ($B$-mode) polarization~\cite{Kamionkowski:1996zd,SZ97,1997PhRvD..55.7368K}. In the standard cosmological model, density perturbations at the last scattering surface produce both temperature and $E$-mode polarization anisotropies. Recent measurements of $E$-mode polarization with the Planck satellite~\cite{Adam:2015rua} are consistent with the standard cosmological model. Density perturbations do not produce primordial $B$-mode polarization at first order. Generating primordial $B$-modes requires sources with parity-odd components, such as gravitational waves~\cite{Crittenden:1993wm,Kamionkowski:1996zd,SZ97}, cosmic birefringence (CB)~\cite{cborigin,1999PhRvL..83.1506L}, primordial magnetic fields (PMFs)~\cite{Kosowsky:1996yc,Seshadri:2000ky,alfvenwaves, alfvenwaves98, tangledB}, or cosmic defects~\cite{Seljak:1997ii}. Here we examine both CB and PMF physics.

Inflation predicts a background of gravitational waves that would produce a primordial $B$-mode signal on degree angular scales. The amplitude of the inflationary $B$-modes is directly related to the energy scale of inflation, a quantity of fundamental importance for anchoring models of the early Universe. On smaller angular scales, gravitational lensing of the CMB by the large-scale structure along the line of sight converts $E$-modes into $B$-modes \cite{1998PhRvD..58b3003Z}. These secondary $B$-modes probe the mass distribution of the Universe and can provide constraints on the sum of neutrino masses~\cite{2015plancklensing,2015planckparameters}. 
In the last two years, a number of experiments started to measure directly the $B$-mode power spectrum, including \pb\ \cite{clbbAPJ}, BICEP2~\cite{bicep2,bicep2planck}, ACTPol~\cite{actbmode}, and SPTpol~\cite{sptpolbmode}.

In addition to probing inflation and the large-scale matter distribution, precision measurements of the CMB $B$-modes promise competitive new tests for a variety of exotic physics. For example, $B$-modes constrain the abundance of cosmic strings and other cosmic defects \cite{moss2014,Lizarraga:2014waa}, supersonic bulk flows~\cite{supersonicflow}, primordial magnetic fields~\cite{Kosowsky:1996yc,Seshadri:2000ky,alfvenwaves, alfvenwaves98, tangledB}, and parity-violating physics~\cite{cborigin,1999PhRvL..83.1506L}. Here we focus on PMFs and parity-violating interactions, both of which lead to birefringence, i.e., a rotation of polarization converting $E$-modes into $B$-modes. $B$-modes generated by parity-violating processes can be compared to those generated by a PMF via Faraday rotation, thus the strength of the parity-violating interaction can be quantified by an equivalent primordial magnetic field level. In addition, the stress energy in the PMF sources vector- and tensor-mode perturbations at the time of last scattering, contributing to the $B$-mode power spectrum~\cite{Seshadri:2000ky,Mack:2001gc,Lewis:2004ef,Paoletti:2008ck}. 

The discrete symmetry groups (charge conjugation, parity and time reversal) play an important role in the standard particle physics model~\cite{0038-5670-34-5-A08}. If both charge and charge plus parity symmetries are not violated then the observed baryon-antibaryon asymmetry cannot exist~\cite{0038-5670-34-5-A08,trodden}.
Cosmological models which also violate these symmetry groups include pseudoscalar models of quintessence~\cite{Frieman:1995pm}, which have the benefit of naturally explaining the smallness of the quintessence field mass and of its coupling to the fields of the standard particle physics model. These models couple the pseudoscalar and electromagnetic fields; the resulting rotation converts $E$-modes into $B$-modes~\cite{Carroll:1998}.

Magnetic fields exist in all gravitationally bound structures in the Universe, from planets and stars to galaxies and galaxy clusters. Explaining the microgauss strength fields observed in galaxies is challenging without a primordial magnetic seed field \cite{RevModPhys.74.775,Durrer:2013pga} coherent over a scale of a few megaparsecs~\cite{2001PhR348163G}. Giving additional impetus to the PMF hypothesis is the claimed detection of magnetic fields in the intergalactic medium~\cite{Neronov_Vovk_2010_science,Tashiro:2013ita}. Candidate mechanisms for the generation of a PMF include inflationary scenarios \cite{Turner:1987bw,Ratra:1991bn} and phase transitions \cite{Vachaspati:1991nm}. Detecting a PMF would lead to important insights into fundamental physics and the early Universe. A recent analysis based on the BICEP2 detection of $B$-mode power~\cite{bicep2} looked for evidence of PMFs in the BICEP2 $B$-mode power spectrum at degree angular scales~\cite{Bonvin:2014xia}. Planck data limits the magnetic field strength smoothed over $1$ Mpc to $B_{1 {\rm Mpc}}<4.4$ nanogauss (nG) at the $95\%$ confidence level~\cite{Ade:2013zuv,2015arXiv150201594P}. Comparable bounds are obtained from Lyman-$\alpha$ spectra~\cite{Kahniashvili:2012dy}. The next generation of CMB polarization experiments promise order of magnitude improvements with the ability to detect sub-nG PMFs~\cite{2009PhRvD..80b3009K,Yadav:2012uz,De:2013dra,Pogosian:2013dya}.

In Sec. II, we use arcminute-scale CMB polarization
data from the \pb\ experiment to constrain
the anisotropic cosmic birefringence power spectrum. An upper limit of an equivalent magnetic field is obtained to interpret this four-point correlation measurement in Sec. III. In Sec. IV, an upper limit on the amplitude of an actual PMF is also constrained by the two-point correlation measurement, i.e., the \pb\ $B$-mode power spectrum. 
The overall structure of these sections is that we discuss parity-violating
physics, i.e., cosmic birefringence, in Secs. II and III, and the primordial
magnetic field in Sec. IV.

\section{Cosmic Birefringence, Faraday Rotation, and the Rotation Angle Estimator}

\subsection{Birefringence and its effect on the CMB}
Cosmic birefringence - the difference in propagation of different polarization states - can rotate CMB polarization and convert $E$-modes to $B$-modes. One proposed source of cosmic birefringence is a coupling between photons and a pseudoscalar field $\phi$. Such couplings arise naturally in modified theories of electromagnetism which include a Chern-Simons term. The Chern-Simons term can appear in pseudoscalar models of quintessence~\cite{prs08,Carroll:1998}, with a Lagrangian

\begin{equation}
{\cal L}= \frac{\phi}{2M}F_{\mu \nu}{\tilde F}^{\mu \nu}\,,
\end{equation}
where $F_{\mu \nu}$ is the electromagnetic field strength tensor, and $\tilde F^{\mu \nu}$ is its dual. The coupling is suppressed by a mass scale $M$. Such an interaction will rotate the linear polarization of the CMB by an angle~\cite{Carroll:1998}  
\begin{equation}
\alpha = \frac{1}{M}\int d\eta \dot{\phi}
\end{equation}
during propagation over an interval in conformal time $\eta$, where $\,\,\dot{\phi}=\partial\phi/\partial\eta$. This rotation of polarization of the CMB creates cosmic birefringence. 

If the spatial average of the field $\left<  \phi \right>$ is not zero, the rotation would produce nonvanishing parity-odd two-point $\langle TB\rangle$ and $\langle EB\rangle$ correlation functions~\cite{1999PhRvL..83.1506L}. Such correlations would imply the existence of a preferred orientation in the Universe and are not normally expected because of the presumed statistical isotropy of cosmological perturbations.

Regardless of the value of $\left<  \phi \right>$, fluctuations in the pseudoscalar field will generate anisotropy in the rotation angle $\alpha$, leading to a spatially varying cosmic birefringence. A statistically isotropic, random $\alpha (\bf{n})$ creates $B$-mode power~\cite{prs08} with an angular dependence determined by the rotation power spectrum. Inhomogeneous cosmic birefringence also correlates the $E$- and $B$-modes, leading to nontrivial four-point correlations. In this paper, we use these four-point correlations to search for anisotropic rotations from CMB maps~\cite{Kamionkowski:2008fp,Yadav_etal_09,2009PhRvD..80b3510G}. The cosmic birefringence constraints in terms of an effective Faraday rotation are discussed at the end of Sec. III.

\subsection{Faraday rotation due to primordial magnetic fields}
The effect of the cosmic birefringence can be described by an equivalent PMF inducing the Faraday rotation. In this section, we exclusively describe the Faraday rotation given by a PMF. The constraints on the actual PMF will be described in Sec. IV.

A PMF embedded in the photon-baryon plasma during recombination will Faraday rotate the plane of polarization of CMB photons, providing another mechanism for cosmic rotation, now with a characteristic frequency dependence. The rotation angle along the line of sight ${\bf n}$ is given by~\cite{Harari:1996ac,Kosowsky:1996yc}
\begin{equation}
\alpha({\bf n}) = \frac{3c^2}{{16 \pi^2 e}} \nu^{-2} 
\int \dot{\tau} \ {\bf B} \cdot d{\bf l} \ ,
\label{theta2}
\end{equation}
where $\dot{\tau}$ is the differential optical depth, $\nu$ is the observed frequency of the radiation, ${\bf B}$ is the comoving magnetic field, $e$ is the electron charge, and $d{\bf l}$ is the comoving length element along the photon trajectory.

A statistically homogeneous, isotropic and Gaussian distributed stochastic magnetic field ${\bf B}(\bf x)$ is characterized by a two-point correlation function in Fourier space~\cite{Mack:2001gc,monin2007statistical} by
\be
\langle B_i ({\bf k} ) B_j ({\bf k}' ) \rangle =
(2\pi )^3 \delta^{(3)}({\bf k} + {\bf k}' )
[ (\delta_{ij} - {\hat k}_i {\hat k}_j) S(k) ],
\label{bcorr}
\ee
where $S(k)$ is the symmetric magnetic field power spectrum and $\hat k_i$ is a normalized component of a wave vector $\bf{k}$. The antisymmetric component describes the helicity of the magnetic field which does not contribute to the Faraday rotation spectrum and has a subdominant contribution to the CMB power spectra, hence we omit the antisymmetric contribution as it is inconsequential for the purpose of this paper~\cite{antisymB1,antisymB2,antisymB3}. The shape of $S(k)$ depends on the mechanism responsible for production of PMF and generally is taken to be a power law up to a certain dissipation scale $k_{\rm{diss}}$. Namely, $S(k) \propto k^n$ for $0<k<k_{\rm diss}$ and zero for $k>k_{\rm diss}$ where $k_{\rm diss}$ depends on the amplitude and the shape of the magnetic field's spectrum and $n$ is the spectral index. For nearly scale-invariant spectra that produce CMB anisotropy, the value of $k_{\rm diss}$ is irrelevant and we take it to be infinite. To quantify the tangling scale of the PMF we smooth its comoving amplitude over a length $\lambda$, obtaining $B_\lambda$. For scale-invariant fields, this quantity is independent of $\lambda$ and is $B_{\rm eff} \equiv \sqrt{8\pi \epsilon_B}$, where $\epsilon_B$ is the total magnetic energy density.

Faraday rotation (FR) happens concurrently with the generation of CMB polarization during recombination. However, Ref.~\cite{Pogosian:2011qv} demonstrated that FR can be applied in a second step (i.e., first produce $E$ modes and then Faraday rotate them by a PMF) without introducing significant errors. In this approximation, the power spectrum of the FR angle can be written as \cite{Pogosian:2011qv}
\begin{eqnarray}
C_L^{\alpha \alpha} &=& {2\over \pi } \int {dk \over k} \Delta^2_M(k) \Big[\frac{L}{2L+1}\mathcal{T}^2_{L-1}(k)\nonumber\\
&+&\frac{L+1}{2L+1}\mathcal{T}^2_{L+1}(k)-\mathcal{T}^{(1)2}_{L}(k)\Big],
\label{c_ell} 
\end{eqnarray}
where $\Delta^2_M(k) \equiv k^3 S(k) [3 c^2 \nu^{-2}/(16 \pi^2 e)]^2$ contains all the physics relevant to PMF, $\mathcal{T}_L(k)$ and $\mathcal{T}^{(1)}_L(k)$  are transfer functions~\cite{Pogosian:2011qv} which are independent of the magnetic field and only depend on the differential optical depth. The discussions in this section will be applied to the Faraday rotation equivalent of the cosmic birefringence measurement in Secs. II and III.

\subsection{Quadratic estimator and previous constraints on rotation power spectrum}

A CMB polarization experiment measures Stokes parameters $Q$ and $U$ at different points on the sky. Anisotropic cosmic birefringence adds a phase factor $e^{\pm 2i\alpha({\bf n})}$ to the underlying primordial CMB polarization.  The Stokes parameters transform as
\begin{equation}
(Q\pm iU)({\bf n})=(\tilde Q\pm i\tilde U)({\bf n})e^{\pm 2i\alpha({\bf n})}, \label{plensed}
\end{equation}
where $\tilde{Q}$ or $\tilde{U}$ denotes the primordial Gaussian CMB polarization map, $Q$ and $U$ are the observed Stokes parameters, and $\alpha({\bf n})$ is the anisotropic rotation field. 
The CMB polarization defined in Eq.~(\ref{plensed}) is rotation invariant and can be decomposed into electric- ($\textit{E}$-) and magneticlike ($\textit{B}$-) modes~\cite{Kamionkowski:1996zd} as
\begin{eqnarray}
 \left[ E\pm i B \right] (\bfl) &=&
        \int  d {\bf n}[Q({\bf n})\pm i U({\bf n})] e^{\mp 2i\phi_{\bf l}} e^{-i {\bf l} \cdot {\bf n}}\,,
\label{EBFields}
\end{eqnarray}
where $\phi_{\bf l}$ is the angular separation between ${\bf n}$ and ${\bf l}$.

Taylor expanding the rotated CMB polarization to first order in the rotation angle reveals that the off-diagonal elements of the two-point correlation functions of $\textit{E}$- and $\textit{B}$-modes are proportional to the rotation field, $\alpha({\bf n})$.
Quadratic estimators take advantage of this feature to measure the anisotropic rotation~\citep{Kamionkowski:2008fp,Yadav_etal_09,2010PhRvD81f3512Y,2009PhRvD..80b3510G}.
The quadratic estimator for CMB polarization is
\begin{equation}
\alpha_{EB}({\bf L})=A_{EB}(L)\int\frac{d^2{\bf l}}{(2\pi)^2}E({\bf l})B({\bf l'})\frac{2\tilde C_l^{EE}\cos2\phi_{{\bf l}{\bf l'}}}{C_l^{EE}C_{l'}^{BB}}\label{EBest},
\end{equation}
where ${\bf l}$, ${\bf l'}$, and ${\bf L}$ are coordinates in Fourier space with  ${\bf L} = {\bf l}+{\bf l'}$. 
The angular separation between ${\bf l}$ and ${\bf l'}$ is $\phi_{{\bf l}{\bf l'}}$, $\tilde C_l^{EE}$ is the theoretical primordial power spectrum, $C_l^{EE}$ and $C_{l}^{BB}$ are $E$- and $B$-mode power spectra that include experimental noise, and $A_{EB}(L)$ is a normalization factor to give an unbiased estimate of the rotation power spectrum~\cite{Yadav_etal_09,Yadav:2012uz}. Note that if the rotation is uniform over the sky, it can be entirely determined by $C_l^{EB}$ and $C_l^{EE}$~\cite{KSY2013}. 

In this work, we focus on the anisotropic rotation rather than the uniform rotation discussed in Refs.~\cite{KSY2013, jonCB, clbbAPJ}. The rotation power spectrum $C_L^{\alpha\alpha}$ is derived from a four-point correlation of $E$ and $B$ via \citep{Kamionkowski:2008fp,Yadav_etal_09,2009PhRvD..80b3510G}
 \begin{eqnarray}
 \langle \alpha_{EB}({\bf L})\alpha^{\ast}_{EB}({\bf L'})\rangle&=&(2\pi)^2\delta({\bf L}-{\bf L'})  (C_L^{\alpha\alpha}+N^{(0)}_{EB}(L)\nonumber\\
&+&\mbox{higher-order terms}),
 \end{eqnarray}
with $N^{(0)}$ being the Gaussian contribution to the four-point function \cite{Yadav:2012uz,Gluscevic:2012me}.

Previous studies have focused on constraining the uniform rotation as well as placing  upper limits on degree-scale rotations~\cite{2012JCAP...02..023G,wmap7rot, 2009PhRvL.102p1302W, jonCB}. Constraints on the anisotropic cosmic birefringence power spectrum have been derived from WMAP-7 data using $\langle TBTB\rangle$ four-point correlations~\cite{Gluscevic:2012me}. In Ref.~\cite{cb2pcorrelation}, the two-point real-space correlation function was used to probe the anisotropic rotation. Both of these analyses limit the anisotropic rotation angle on large scales to be less than a few degrees.

\section{Bounds on anisotropic rotation from POLARBEAR}
\subsection{Data analysis}
The \pb\ telescope is located in the Atacama Desert in northern Chile and observes in a band centered at 148 GHz with a beam size
of $3.5'$ full width at half maximum. This analysis uses data on three regions selected for their low dust emission, hereafter referred to as RA4.5, RA12, and RA23 based on their right ascensions~\cite{clbbAPJ}. The total area of the three patches is 25 square degrees, and the patches were observed by the \pb\ experiment during June 2012 to June 2013. This data is referred to as the first-season \pb\ data.

The time ordered data (TOD) from the detectors are filtered and coadded into maps as described by Ref.~\cite{clbbAPJ}.
We first flag and remove data affected by spurious instrumental or environmental effects.
The TOD are bandpass filtered with the upper band edge set by a low-pass filter and the lower band edge set by the subtraction of a first-order polynomial from each constant-elevation, constant-velocity subscan. 

A ground template, fixed in azimuth, is also removed. 
Bright radio sources are masked before removing the ground template and polynomial. 
Each pixel consists of two bolometers sensitive to orthogonal polarization; data from these two bolometers are summed and differenced to derive temperature and polarization TOD from each pixel. The TOD are then coadded with inverse variance weighting into maps according to a weight estimated from the average power spectral density between 1 and 3 Hz of the filtered TOD.

We construct an apodization window from the smoothed inverse variance weight map. Pixels with an apodization window value below 1\% of the peak value are set to zero, as are pixels within $3'$ of bright sources in the Australia Telescope 20 GHz Survey \cite{AT20G}. 
\textit{Q} and \textit{U} maps are transformed to \textit{E} and \textit{B} maps using the pure-B transform ~\cite{PureEstimator_Smith2006}. The instrument polarization angle is calibrated using the patch-combined $C_l^{EB}$ power spectrum~\cite{KSY2013, clbbAPJ}, so the monopole contribution to the anisotropic rotation is removed. 

We reconstruct the rotation field by applying the estimator in Eq. (\ref{EBest}) to the coadded \pb\ maps for $l,l^\prime \in \{500,2700\}$. The reconstructed rotation power spectrum is calculated as follows: 
\begin{equation}
C_L^{\alpha\alpha}=(\langle \alpha({\bf L})\alpha^{\ast}({\bf L})\rangle-N_L^{(0)}) / T_L,
\end{equation}
with both the Gaussian bias $N_L^{(0)}$ and the transfer function $T_L$ calculated using simulations. The mean estimated rotation is subtracted from the reconstructions and the realization-dependent Gaussian bias is subtracted for the final results~\cite{2014AA571A17P,2014arXiv1412.4760S}. 

We create simulated map realizations of the theoretical spectra calculated by CAMB~\cite{camb}. For the simulated rotation maps, we assume a scale-invariant power spectrum $L(L+1)C_L^{\alpha\alpha}/2\pi=10^{-4}{\ }\rm{rad}^2$ (0.33 $\rm{deg}^2$). 
In the rotation simulation, map pixels are multiplied by a phase factor following Eq. (\ref{plensed}) to obtain rotated polarization maps. We convolve each realization by the measured beam profile and a transfer function that accounts for the filtering on the time stream, and add noise based on the observed noise levels in the polarization maps. The finite area of the \pb\ fields results in a window function that couples to large-scale modes, biasing them at low $L$. We correct this bias by calculating a transfer function from the ratio of the averaged reconstructed rotation power spectrum to the known input for $L<400$. We validate the rotation reconstruction by correlating the estimated rotation maps with the input maps whose rotation power spectra are known. All the spectra for all patches agree with the input rotation power spectra.

\subsection{Systematic errors and null tests}
Systematic errors can generate spurious signals which might mimic the signals we want to probe. Possible sources of systematic error in the \pb\ $B$-mode power spectrum have been studied extensively by Ref.~\cite{clbbAPJ}. In this section, we extend the foreground modeling of Ref.~\cite{clbbAPJ} to the cosmic birefringence signal and present a new systematic null test specific to the birefringence analysis.

Astrophysical foregrounds might affect measurements of the anisotropic cosmic birefringence. We test the impact of foregrounds in four ways. First, we generate Gaussian realizations of the foreground emission due to galactic dust and synchrotron, and radio and dusty galaxies. The templates and amplitudes for each term are taken from the default foreground model presented by Ref.~\cite{clbbAPJ}. The foreground realizations are added to simulated CMB maps, and the rotation power spectra are estimated. We find adding foregrounds does not bias the result, but does negligibly increase the uncertainty by 0.6\%.

A potential concern about this treatment is that the radio galaxies, which are effectively unresolved point sources, should be drawn from a Poisson instead of Gaussian distribution. We therefore create a set of Poisson realizations drawn from the empirically determined number counts $dN/dS\propto S^{\alpha}$, where $\alpha=-2.15$~\cite{cmbpol} and $S$ is the source flux. We limit the distribution to fluxes with $1{\ }\rm{mJy}<$$S$$<25{\ }\rm{mJy}$. The upper limit is set by the point source detection threshold and the lower limit is chosen to be sufficiently small that it has no effect on the simulated power. The equivalent power of the unresolved point sources is $7{\ }\mu{\rm K}^2$ at $l=3000$. We assume a 5\% polarization fraction and random polarization angles. We propagate the polarization maps of the unresolved point sources and find the simulated contamination to be negligible.

Finally, we perform two tests to quantify the worst-case impact of polarized galactic dust. First, we change the dust polarization spectrum from the default model in Ref.~\cite{clbbAPJ} to the higher, empirical model of Ref.~\cite{clbbAPJ}. As before, we generate Gaussian realizations of the dust polarization in our observing patches. We find that the higher empirical dust model from Ref.~\cite{clbbAPJ} leads to a larger, but still negligible increase in the uncertainty of 1.2\%. Second, we test whether the non-Gaussianity of the polarized dust anisotropy is important. For this, we use the dust template from the Planck Sky Model~\cite{psm} and extract anisotropic dust maps for all \pb\ patches. The polarized dust maps are multiplied by a factor of 2 since the Planck dust study indicates a larger polarized
fraction~\cite{2014arXiv1405.0871P, PIPXXX}. The rotation power spectrum of these dust maps shifts the best-fit result by $0.1\sigma$. As we only have one template, it is impossible to determine if this is a bias or scatter. Both tests argue that the polarized dust has only a small effect on
the measured rotation power spectra. We conservatively weaken the upper limit on the anisotropic birefringence signal by $0.2 \sigma$ to account for foregrounds.

\begin{figure}
\rotatebox{0}{\includegraphics[width=9cm, height=6cm]{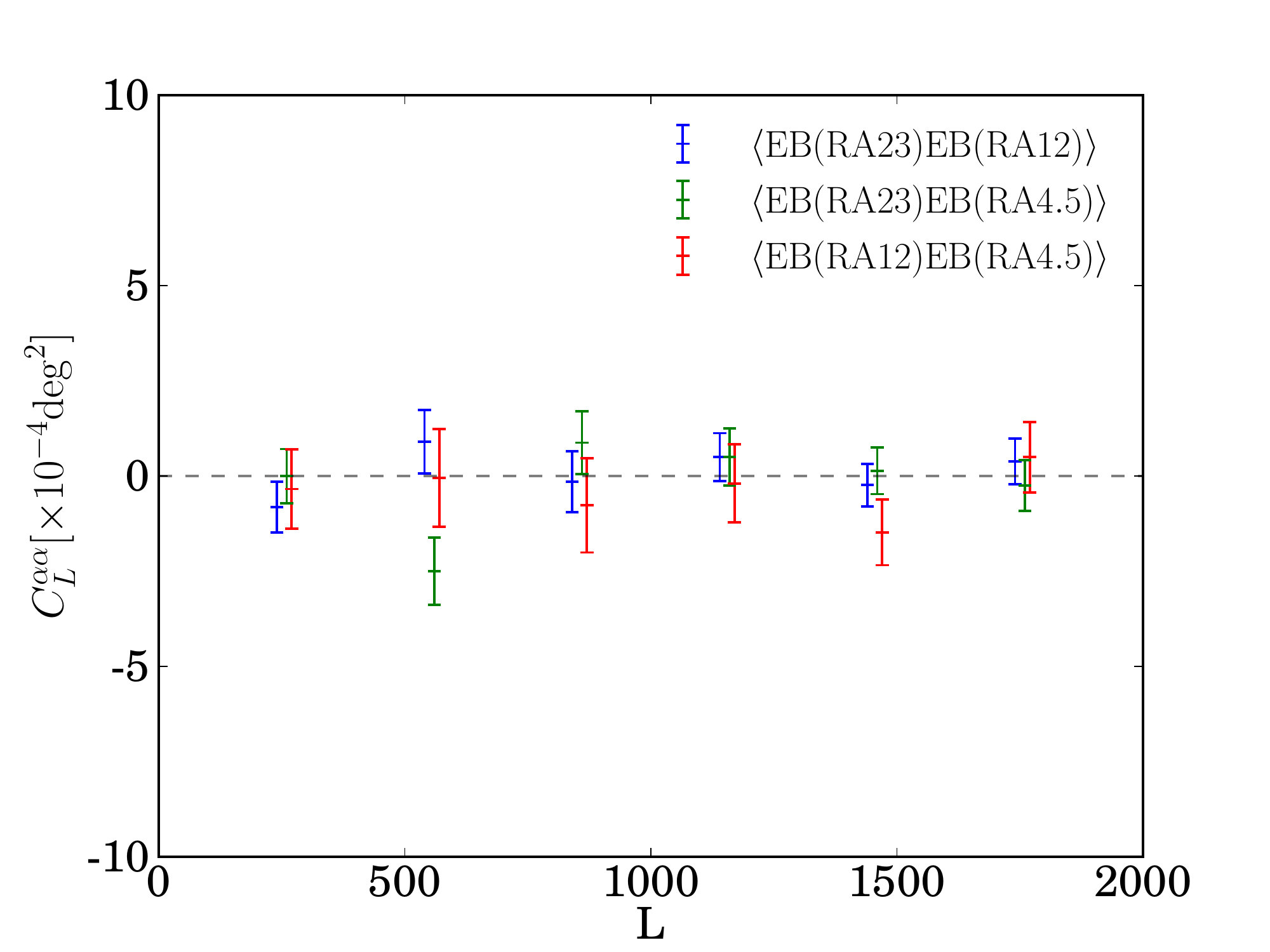}}
\caption{Swap-patch rotation power spectra are shown for each of the three patches. The power spectra are calculated from the rotation fields on different patches and the legend indicates a specific combination. The data show no evidence for systematic contamination. }\label{swap}
\end{figure}

Rotation fields for different patches should be uncorrelated. We use this fact to test for potential contamination via the ``swap-patch" null test. We define a swap-patch rotation power spectra $C_L^{\alpha\alpha, \rm{null}}=\langle \alpha_{{\rm patch}{\ }1}({\bf L})\alpha_{\rm{patch}{\ }2}^{\ast}({\bf L})\rangle$~\cite{Das2011}. In the absence of contamination, the swap-patch power spectrum should be consistent with zero. The results are shown in Fig.~\ref{swap}. The probability-to-exceed values (PTEs) are $64\%, 10\%, 71\%$ for the three combinations $\langle\alpha_{\rm{RA23}}({\bf L})\alpha_{\rm{RA12}}^{\ast}({\bf L})\rangle$, $\langle\alpha_{\rm{RA23}}({\bf L})\alpha_{\rm{RA4.5}}^{\ast}({\bf L})\rangle$ and $\langle\alpha_{\rm{RA12}}({\bf L})\alpha_{\rm{RA4.5}}^{\ast}({\bf L})\rangle$. We find no evidence of contamination.

\begin{figure}
\rotatebox{-0}{\includegraphics[width=9cm, height=6cm]{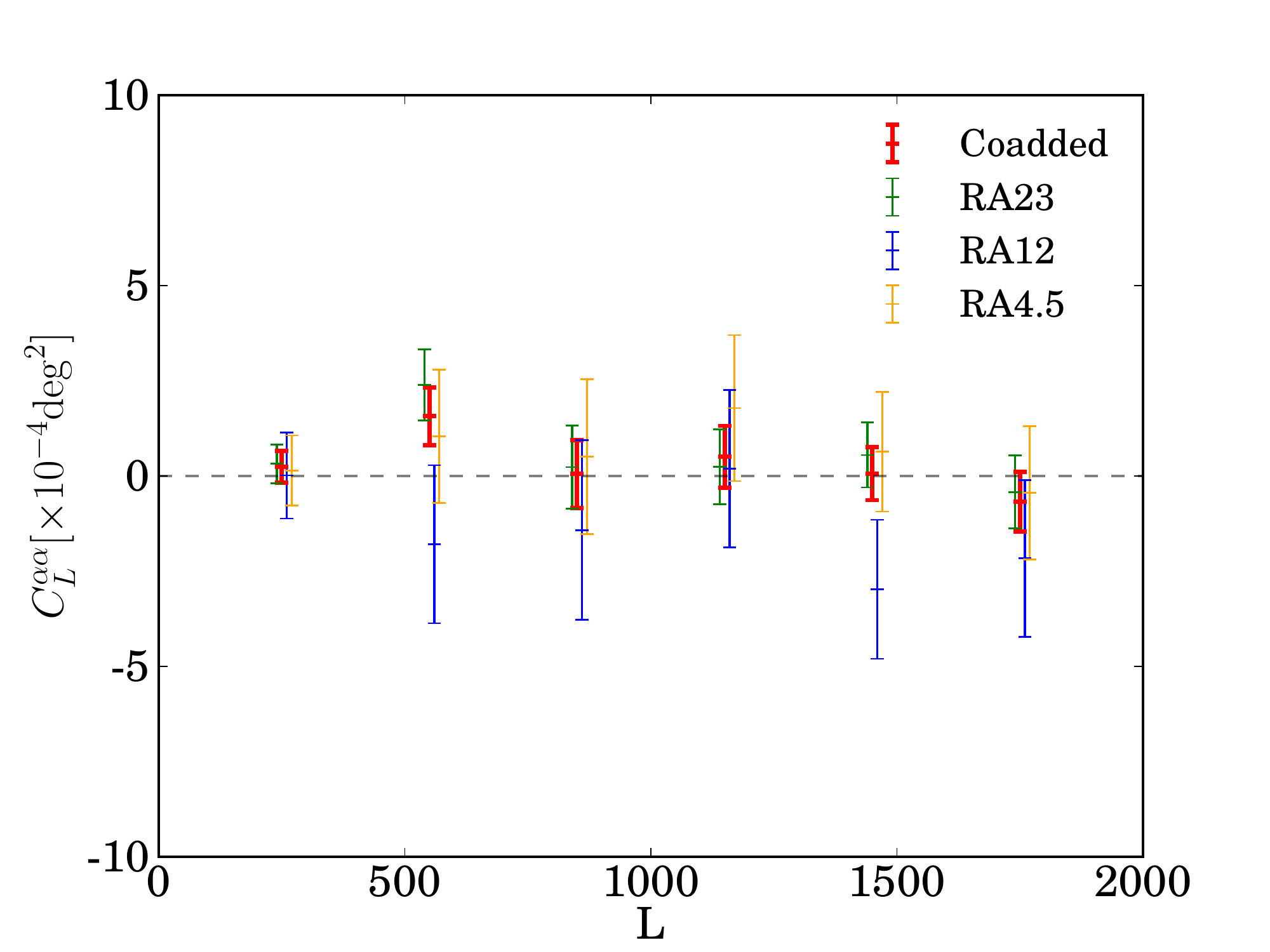}}
\caption{The anisotropic cosmic rotation power spectra from \pb\ 's first-season data in three patches. The spectrum of an individual patch is indicated by the green (RA23), blue (RA12) and orange (RA4.5) colors. The coadded (red) power spectrum is consistent with zero. }
\label{claa}
\end{figure}

\begin{figure}
\rotatebox{0}{\includegraphics[width=8cm, height=6cm]{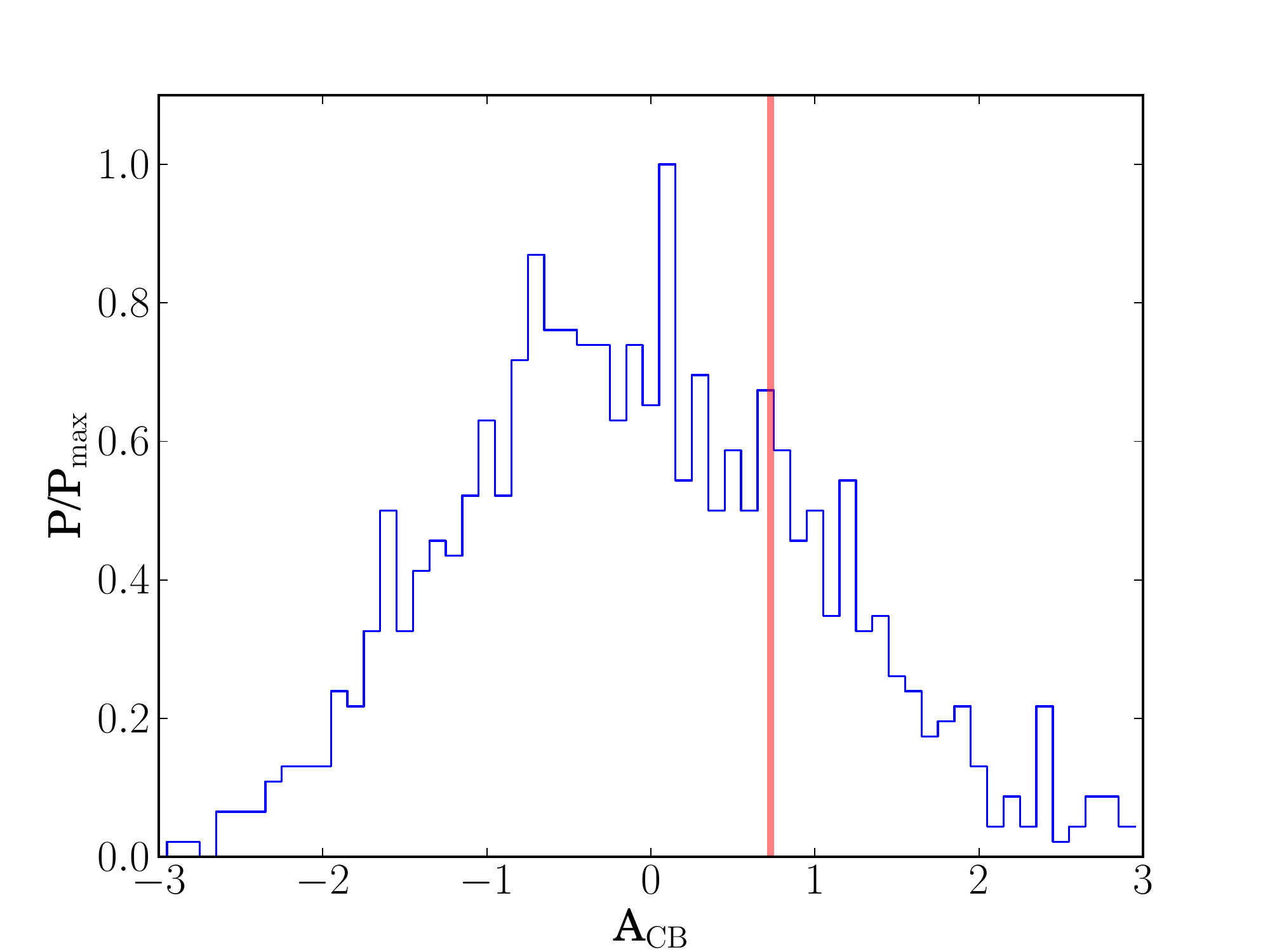}}
\caption{The blue histogram shows the distribution of the amplitude $A_{\rm{CB}}$ from null signal simulations. The red vertical line corresponds to the best-fit amplitude that minimizes the $\chi^2$ in Eq.~(\ref{chi2}).}\label{claaAmp}
\end{figure}

\subsection{Results}
In Fig.~\ref{claa}, we show the anisotropic cosmic birefringence power spectrum reconstructed using the quadratic estimator from data in three different patches, as well as the coadded spectrum.  The measurement is consistent with zero and we do not detect any anisotropic rotation signal from \pb\ data. 

A scale-invariant rotation power spectrum is particularly interesting because it could result from inflationary fluctuations
of a massless pseudoscalar \cite{prs08}. For a scale-invariant rotation field,  $L(L+1)C_L/2\pi \approx {\rm const}$. We define a dimensionless amplitude parameter ${A}_{\rm{CB}}$ as a factor relating an arbitrary scale-invariant spectrum to a reference spectrum, $C_L^{\rm ref}$, for which $L(L+1)C_L^{\rm ref}/2\pi = 10^{-4}{\ }\rm{rad}^2$ (0.33 $\rm{deg}^2$). In the WMAP analysis~\cite{Gluscevic:2012me}, a scale-invariant power spectrum with an amplitude $6\times 10^{-3}{\ }\rm{rad}^2$ (21 $\rm{deg}^2$) is adopted. The best-fit amplitude of the scale-invariant anisotropic rotation power spectrum corresponds to the minimum of 
\be
\chi^2({A}_{\rm{CB}})=\sum_{bb'}(\hat C^{\rm{obs}}_b-{A}_{\rm{CB}}C^{\rm{ref}}_{b})M^{-1}_{bb'}(\hat C^{\rm{obs}}_{b'}-{A}_{\rm{CB}}C^{\rm{ref}}_{b'})
\label{chi2}
\ee
where $b$ is the index of the rotation band power and $\hat C_b^{\rm{obs}}$ is the measured spectrum in band $b$. The covariance matrix $M_{bb'}$ is calculated from simulations with no cosmic birefringence signal. The posterior distribution is shown in Fig.~\ref{claaAmp}.

An upper limit on the amplitude of the rotation spectrum can be interpreted as a bound on the magnitude of FR and the magnetic field spectrum. A scale-invariant PMF results in a scale-invariant FR spectrum \cite{Pogosian:2011qv}. At the \pb\ frequency $\nu=148$ GHz, the measured 95\% confidence limit $A_{\rm CB} < 3.1$ translates into a four-point correlation bound on the strength of an equivalent PMF: $B_{1\rm{Mpc}} < 90$ nG, according to the relation $B_{\rm{1Mpc}}=(2.1\times10^2{\ }\rm{nG})(\nu/30{\ }\rm{GHz})^2$$\sqrt{L(L+1)C_L^{\alpha\alpha}/2\pi}$~\cite{De:2013dra,Pogosian:2013dya}. Including estimates for known systematic errors, this limit becomes $B_{1\rm{Mpc}} < 93$ nG. Our constraint from the cosmic birefringence power spectrum is roughly 15 times lower than the recent 95\% confidence level limit of $B_{1\rm{Mpc}} < 1380$ nG inferred from constraining the contribution of Faraday rotation to the Planck polarization power spectra~\cite{2015arXiv150201594P}. Also, compared to the WMAP cosmic birefringence measurement~\cite{Gluscevic:2012me}, the amplitude of the rotation power spectrum from \pb\ is roughly $60$ times smaller.

\section{Constraints on primordial magnetic fields from the $B$-mode power spectrum}
\label{sec:bb}
\begin{figure}[tbp]
\includegraphics[width=1.\columnwidth,angle=0]{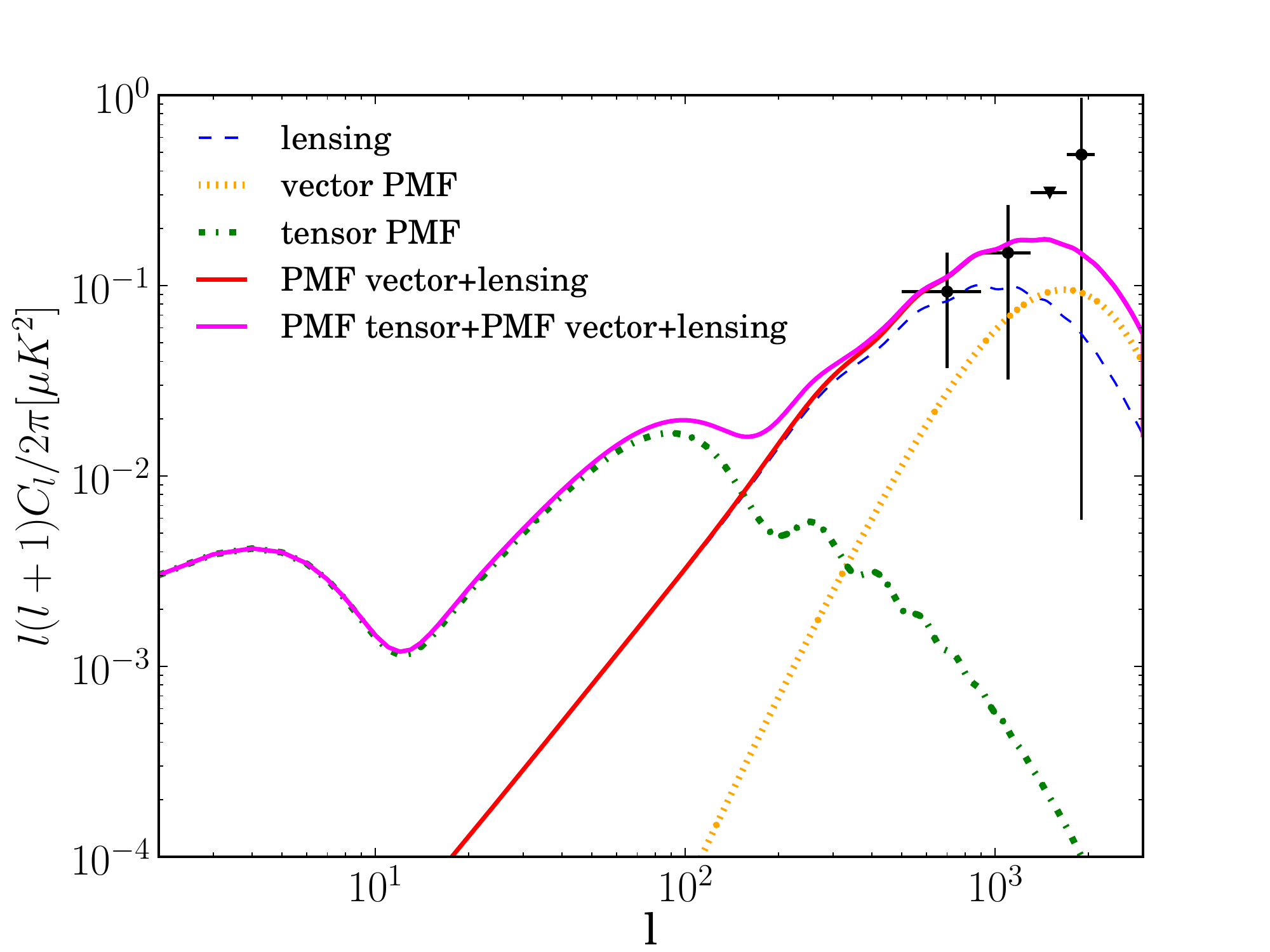}
\caption{A representative $B$-mode polarization power spectrum sourced by a scale-invariant PMF.  Shown are the passive tensor mode (green), the compensated vector mode (orange), the gravitational lensing contribution (blue) and the combinations of the lensing and vector $B$ modes (red) and all three components (magenta).  The PMF contribution is based on $B_{\rm{1Mpc}}=2.5$ \rm{nG}, $n=-2.9$, $a_\nu/a_{\rm{PMF}}=10^{9}$. The data points are from the \pb\ first-season $B$-mode power spectrum. The third point is the 95\% upper limit assuming the band power is positive.}
\label{fig:BBpmf}
\end{figure}

The stress energy in the PMF sources vector- and tensor-mode perturbations in the metric leading to a frequency independent contribution to the CMB's $B$-mode polarization \cite{Lewis:2004ef}. This contribution is in addition to the frequency dependent FR signal discussed earlier. There are two potentially observable frequency independent contributions to the $B$-mode spectrum from a nearly scale-invariant PMF \cite{2010PhRvD..81d3517S,Bonvin:2014xia}. One comes from the passive, or uncompensated tensor mode, which is generated by the PMF before neutrino decoupling. As shown with the dash-dotted green line in Fig.~\ref{fig:BBpmf}, the spectrum of this component is practically indistinguishable from the inflationary gravity wave signal. The amplitude of the tensor contribution is proportional to $B^4_{1\rm{Mpc}} [\ln(a_\nu / a_{\rm{PMF}})]^2$, where $a_\nu$ is the scale factor at neutrino decoupling and $a_{\rm{PMF}}$ is the scale factor at which PMF was generated. The passive tensor mode is not constrained by the existing \pb\ analysis, which only probes $l > 500$  \cite{clbbAPJ}. However, future measurements of $C_l^{BB} $ at $l<100$ will probe the tensor contribution, although it will likely be degenerate with primordial gravitational waves. 

The PMF vector modes are more directly relevant to the current \pb\ data as shown by the dotted orange line in Fig.~\ref{fig:BBpmf}. The $B$-mode power spectrum generated by a scale-invariant PMF peaks around $l\sim 1700$, with the peak power given by 
\be
 {l(l+1)C_l^{BB} \over 2\pi}\Big{|}_{l \sim 1700} \sim 2.5 \times 10^{-3} \left( B_{\rm 1Mpc} \over {\rm nG} \right)^4  \mu {\rm K}^2 \ .
 \label{l2000}
\ee
The vector-mode contribution is independent of $a_{\rm{PMF}}$. 

Therefore, the PMF $B$-mode power spectrum can be characterized by three parameters: the PMF amplitude $B_{\rm{1Mpc}}$, the epoch of PMF generation $\beta=\ln{(a_{\nu}/a_{\rm{PMF}})}$, and the PMF spectral index $n$, where we note that the parameter $\beta$ only affects the tensor mode. In what follows, we use the \pb\ $B$-mode power spectrum~\cite{clbbAPJ} to derive constraints on $B_{1\rm{Mpc}}$, marginalizing over the other parameters.

\subsection{Data analysis}

Our theoretical $B$-mode model consists of lensing and the PMF vector $B$-modes. \pb\ data measured the $B$-mode power spectrum at 148 GHz~\cite{clbbAPJ}. We use the published \pb\ $B$-mode window functions, band power and band variances to construct the likelihood function. We assume a Gaussian likelihood for the \pb\ data and adopt the following priors on the PMF parameters: $0<B_{\rm{1Mpc}}<10{\ }\rm{nG}$, $-2.9<n<-1.5$ and $0<\beta<39$. A larger prior upper limit on $B_{\rm{1Mpc}}$ is not necessary because constraints obtained in this analysis are well below this bound. The upper prior on $n$ is chosen because for high $n$, or ``bluer" PMF spectra, most of the PMF energy is concentrated on small scales, with only negligible power on scales above 1 Mpc that are of relevance to our data. Thus, extending the range of $n$ would make no difference for our constraints, unless we allow for an extremely strong PMF, which is ruled out. On the other hand, the spectral index has to be larger than $-3$ to avoid the divergence of the PMF power spectrum. We take into account the systematic contamination of the \pb\ $B$-mode power spectrum considered in Ref.~\cite{clbbAPJ} and investigate how the systematic uncertainties can potentially affect the PMF constraints. 

\subsection{Results}

\begin{figure}[tbp]
\includegraphics[width=1.\columnwidth,angle=0]{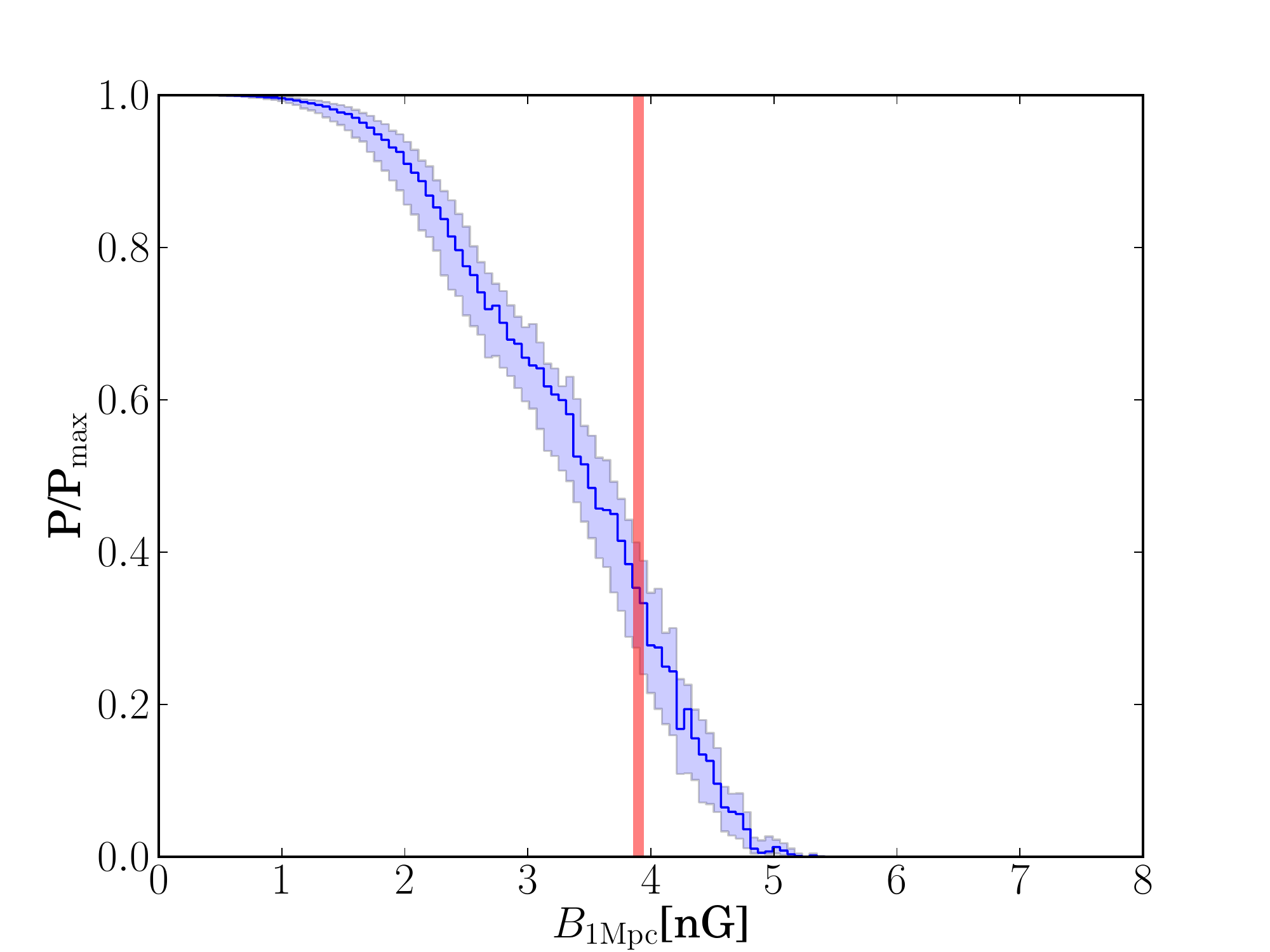}
\caption{Posterior distribution function of amplitude $B_{1\rm{Mpc}}$ of primordial magnetic field using \pb\ first-season $C_l^{\rm{BB}}$ measurement. The vertical line indicates the $95\%$ confidence level upper limit at $B_{1\rm{Mpc}}<$ \pmfupper\ nG. The shaded area is the variations introduced by both the systematic and multiplicative effects.}
\label{fig:posteriorPB}
\end{figure}

In Fig.~\ref{fig:posteriorPB} we show the marginalized posterior distribution function (PDF) of the PMF amplitude $B_{1\rm{Mpc}}$. We take advantage of the detailed study of systematic uncertainties affecting the $B$-mode power spectrum in Ref.~\cite{clbbAPJ} to investigate the effects on the PMF constraints. The PDF without systematics is in blue and the shaded area indicates the shift of the PDF when all known sources of systematic error are included. The likelihood function peaks at $B_{1\rm{Mpc}}$\,=\,0, thus only the upper bound can be derived. It is determined by integrating the area and the vertical red line shows the 95\% bound of \pmfupper\ \rm{nG}; systematic errors have a negligible impact of $\sim$ 5\%. We have examined the posterior distribution of the spectral index $n$, and find as expected that the \pb\ data do not constrain the spectral index. The PMF amplitude constraint from the first-season \pb\ $B$-mode power spectrum alone is comparable to the Planck 2015 limit of $B_{1\rm{Mpc}}< 4.4$ \rm{nG} at 95\% confidence level, where the Planck results include both temperature and polarization information~\cite{2015arXiv150201594P}. 

We have assumed a flat prior on $B_{1\rm{Mpc}}$ for the constraint in Fig.~\ref{fig:posteriorPB}, following the usual convention in the literature.
Note that, as expected with a limit, the prior choice has a substantial effect on the resulting posterior and inferred limits.
We investigate a uniform prior in the space of the observationally constrained quantity: $B^4_{1\rm{Mpc}}$.
The 95\% C.L. upper limit for this case increases somewhat to 4.5 \rm{nG}.
Another prior, the Jeffrey's prior which is uniform in $\log_{10}[B_{1\rm{Mpc}}/\rm{nG}]$, is frequently used for parameters whose magnitude is unknown.
However the posterior for the Jeffrey's prior diverges in this case due to the lack of a reliable lower bound on $B_{1\rm{Mpc}}$, a conclusion also reached in Ref.~\cite{logprior}.
Alternatively, we can examine for what value of $B_{1\rm{Mpc}}$ increases the $\chi^2$ by 4 (analagous to 2\,$\sigma$) relative to $B_{1\rm{Mpc}}$\,=\,0; we find this occurs at $B_{1\rm{Mpc}}$\,=\,4.4\,\rm{nG}. Based on these tests, high PMF amplitudes ($B_{1\rm{Mpc}} >$\,4.5 \rm{nG}) are  disfavored at the 95\% C.L. by the \pb\ $B$-mode power spectrum measurement.

\section{Conclusions}

We constrain the anisotropic rotation power spectrum from the first-season of \pb\ data using the four-point correlations of the CMB polarization. The amplitude of this spectrum is consistent with zero and we do not detect anisotropic cosmic birefringence effects. The amplitude of an equivalent PMF interpreted from the anisotropic rotation power spectrum is less than 90 \rm{nG} (93 \rm{nG}) at the $95\%$ confidence level from the four-point correlation functions, without (with) systematic uncertainties included. 

We also use the first-season \pb\ $B$-mode power spectrum to constrain the magnetically induced vector-mode contribution to $B$-modes. We find that the PMF amplitude from the two-point correlation functions is less than \pmfupper\ \rm{nG} at the $95\%$ confidence level, assuming a flat prior on the PMF amplitude. This limit increases to 4.5 \rm{nG} if we instead adopt a uniform prior on the PMF-sourced $B$-mode power. Neither the anisotropic rotation power spectrum nor the PMF constraints show evidence for significant systematic errors. 

Anisotropic cosmic birefringence directly probes the $B$-mode contribution created by the parity-violating physics as measured by the four-point correlations of the CMB polarization at different angular scales. On the other hand, the two-point correlation function, i.e., the $B$-mode power spectrum measures all curl-like polarization patterns which could be introduced by different sources, such as primordial tensor perturbations, gravitational lensing effects, PMF vector (and tensor) perturbations, Faraday rotation, and parity-violating interactions. Thus $B$-mode power spectra can provide upper limits on the amplitude of PMF, and the four-point correlation measurement would potentially distinguish the rotation mechanisms of the parity-violating physics from PMF with the upcoming multifrequency CMB experiments. For example, future Planck polarization data could be used to measure anisotropic rotation power spectrum over the entire sky and possibly achieve a lower FR upper limit, but our results will be complementary since the two experiments probe the cosmic birefringence effects on different angular scales. 

CMB polarization data will be complemented by other measurements as well, such as observations of $\gamma$-ray emission from blazars~\cite{2011MNRAS.414.3566T,2011A&A529A.144T,2011ApJ727L4D}. Together, these windows on the fundamental physics of the early Universe will help characterize the nature of parity-violating physics and primordial magnetism.\\\\

\acknowledgments
Calculations were performed on the Gordon supercomputer operated for
the Extreme Science and Engineering Discovery Environment by the San
Diego Supercomputer Center and the Edison supercomputer by the National Energy Research Scientific Computing, supported by the Department of Energy under Contract No. DE-AC02-05CH11231.
The \Pb{} project is funded by the National Science Foundation under Grants No. AST-0618398 and No. AST-1212230. 
The KEK authors were supported by MEXT KAKENHI Grants No. 21111002 and No. 26220709, and acknowledge support from KEK Cryogenics Science Center. This work was supported by the JSPS Core-to-Core Program (A. Advanced Research Networks).
The McGill authors acknowledge funding from the Natural Sciences and Engineering Research Council and Canadian Institute for Advanced Research. L.P. and Y.L. are supported by a Discovery Grant from NSERC. The James Ax Observatory operates in the Parque Astron\'{o}mico
Atacama in Northern Chile under the auspices of the Comisi\'{o}n Nacional de Investigaci\'{o}n Cient\'{i}fica y Tecnol\'{o}gica de Chile (CONICYT). K.A. acknowledges support from the Simons Foundation. C.F. acknowledges support from NSF Grant No. AST-1313319 and the Ax Center for Experimental Cosmology. C.R. acknowledges support from the University of Melbourne. C.B., G.F., and G.P. acknowledge partial support from the INDARK INFN Network. D.B. is supported by a NSF Astronomy and Astrophysics Postdoctoral Fellowship under Grant No. AST-1501422. We are grateful to Marc Kamionkowski and Vera Gluscevic for the insights and suggestions that helped inspire this work. 

\bibliography{cbpmf}

\end{document}